\documentclass[aps,pra,reprint,superscriptaddress]{revtex4-2}
\usepackage{amsmath,amsfonts,amssymb}
\usepackage{xcolor,graphicx}
\usepackage{tikz}
\usepackage{color}
\usepackage[pdftex]{hyperref} 
\usepackage{braket}
\usepackage{dcolumn}
\usepackage{bm}
\hypersetup{
	colorlinks = true,
	linkcolor = blue,
	anchorcolor = blue,
	citecolor = blue,
	filecolor = blue,
	urlcolor = blue}
\begin{document}
\title{Exact solution for the interaction of two decaying quantized fields}
\author{L.~Hernández-Sánchez}
\affiliation{Instituto Nacional de Astrofísica Óptica y Electrónica, Calle Luis Enrique Erro No. 1\\ Santa María Tonantzintla, Pue., 72840, Mexico}
\author{I. Ramos-Prieto}
\email[e-mail: ]{iran@inaoep.mx}
\affiliation{Instituto Nacional de Astrofísica Óptica y Electrónica, Calle Luis Enrique Erro No. 1\\ Santa María Tonantzintla, Pue., 72840, Mexico}
\author{F. Soto-Eguibar}
\affiliation{Instituto Nacional de Astrofísica Óptica y Electrónica, Calle Luis Enrique Erro No. 1\\ Santa María Tonantzintla, Pue., 72840, Mexico}
\author{H. M. Moya-Cessa}
\affiliation{Instituto Nacional de Astrofísica Óptica y Electrónica, Calle Luis Enrique Erro No. 1\\ Santa María Tonantzintla, Pue., 72840, Mexico}
\email[e-mail: ]{hmmc@inaoep.mx}
\begin{abstract}
We show that the Markovian dynamics of two coupled harmonic oscillators may be analyzed using a Schrödinger equation and an effective non-Hermitian Hamiltonian. This may be achieved by a non-unitary transformation that involves superoperators; such transformation enables the removal of quantum jump superoperators, that allows us to rewrite the Lindblad master equation in terms of a von Neumann-like equation with an effective non-Hermitian Hamiltonian. This may be generalized to an arbitrary number of interacting fields. Finally, by applying an extra non-unitary transformation, we may diagonalize the effective non-Hermitian Hamiltonian to obtain the evolution of any input state in a fully quantum domain.
\end{abstract}
\date{\today}
\maketitle
When a quantum system is influenced by its environment, one of the main tools to describe the Markovian dynamics is the Lindblad master equation~\cite{Breuer_Book}. It is instructive to rewrite the Lindblad master equation in two parts: on the one hand, the terms that preserve the number of excitations in an effective non-Hermitian Hamiltonian, and on the other hand, the remaining terms that describe the quantum jumps of the system (de-excitations); understanding and studying Markovian dynamics in this way is intimately related to the quantum trajectories technique or the Monte-Carlo wave-function method~\cite{Carmichael_Book,Dum_1992,Molmer_1993}. Nevertheless, even if the non-unitary effective part of the equation is solvable, the full solution of the Linblad master equation is not a trivial task~\cite{Prosen_2012,Torres_2014,Minganti_2019,Arkhipov_2020,Minganti_2020,Teuber_2020}. Despite this, in non-Hermitian or semi-classical approaches, interest in such non-Hermitian systems has vastly grown due to their unusual properties, and a hallmark of these systems are the concepts of parity-time~($\mathcal{PT}$)~symmetry and exceptional points (EPs)~\cite{Bender_1998,ElGanainy_2007,Ruter_2010,Heiss_2012,Kato,MohammadAli_2019,Quiroz_2019,Tschernig_2022}. In addition to the above, in a fully quantum domain, where the light behaves non-classically, multi-particle quantum interference phenomena and non-classical states arise \cite{Lai_1991,Klauck_2019,Longhi_2020}. In this sense, and although the quantum jump superoperators are suppressed in the non-Hermitian Hamiltonian formalism, from our perspective there is a question that must be answered: Is it possible to establish a direct equivalence with the Markovian dynamics through some transformation?; this point represents the main motivation of the present work.

On the other hand, since the paraxial equation and the Schrödinger equation are isomorphic under certain conditions (see, for instance,~\cite{Longhi_2009} and references
therein), photonic structures or evanescently coupled waveguides are the main platforms for directly observing non-unitary evolution~\cite{ElGanainy_2007,Guo_2009, Ruter_2010,Ornigotti_2014,Zhang_2016}. Significantly, non-Hermitian systems exhibiting $\mathcal{PT}$-symmetry and exceptional points have generated a wide range of applications, e.g., loss-induced transparency, laser mode control, optical detection and unidirectional invisibility, just to name a few \cite{Guo_2009,Lin_2011,Feng_2017,ElGanainy_2018,MohammadAli_2019,Ozdemir_2019}. Recently, two representative examples of two-photon quantum interference have been considered in photonic systems with passive losses: Hong-Ou-Mandel dip in a passive $\mathcal{PT}$-symmetric optical directional coupler~\cite{Hong-Ou-Mandel,Klauck_2019}, and a phase transition at the coalescence point~\cite{Longhi_2020}. However, although dissipation process can be described by coupling one or a set of waveguides to one or more reservoirs at zero temperature~\cite{Klauck_2019}, one hopes to have greater control, a priory, over the photonic platform if the quantum jump superoperators are judiciously removed. In fact, removing these superoperators under some transformation leads to the non-Hermitian Hamiltonian formalism.

\begin{figure}[ht!]
\begin{center}
\includegraphics{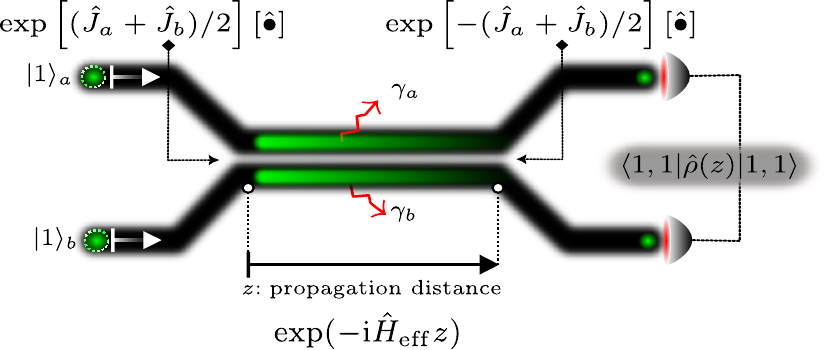}
\caption{Schematic representation of two coupled waveguides with decay rate $\gamma_a$ and $\gamma_b$, respectively. At $z=0$, the photonic state is transformed via $\hat{\varrho}(0)=\exp\left[(\hat{J}_a+\hat{J}_b)/2\right]\hat{\rho}(0)$, to then cross a passive $\mathcal{PT}$-symmetric optical system~\cite{Ornigotti_2014}, whose evolution as a function of the propagation distance $z$ is determined by $\hat{\varrho}(z)=\hat{U}(z)\hat{\varrho}(0)\hat{U}^{-1}(z)$. Finally, before detection, the photonic state is transformed via $\exp\left[-(\hat{J}_a+\hat{J}_b)/2\right]\hat{\varrho}(z)$, in order to obtain the density matrix $\hat{\rho}(z)$, \eqref{rho}.}
\label{Fig_1}
\end{center}
\end{figure}

In this contribution, we consider two quantized fields experiencing (Markovian) losses. These systems have been used to describe other physical systems of bosonic nature, e.g., states of light traveling along two evanescently coupled waveguides~\cite{Lai_1991,Klauck_2019,Longhi_2020}. By means of two transformations, we demonstrate the following: a) the non-unitary dynamics, governed by the Lindblad master equation and the von Neumann-like equation, with an effective non-Hermitian Hamiltonian, are equivalent by removing the quantum jump superoperators by means of the transformation $e^{\pm(\hat{J}_a+\hat{J}_b)/2}\hat{\rho}$; b) we diagonalize the effective non-Hermitian Hamiltonian to obtain the evolution of any input state in a fully quantum domain. The above are the main contributions of this work, because any non-classical state that is constrained to Markovian dynamics, can be equivalently described in terms of light state crossing non-Hermitian systems~(e.g., waveguides or directional coupler) with~(experimentally feasible) passive losses, provided that the input and output states are transformed by means of~$\exp\left[\pm(\hat{J}_a+\hat{J}_b)/2\right]\left[\hat{\bullet}\right]$ (see Fig.~\ref{Fig_1}). We would like to stress that such transformation may be easily generalized to an arbitrary number of interacting and decaying fields.

Let us consider the interaction between two quantized fields subject to Markovian losses, determined by the loss rates $\gamma_a$ and $\gamma_b$, respectively. The Markovian dynamics in the interaction picture for the reduced density matrix $\hat{\rho}$ is governed by the Lindblad master equation~\cite{Breuer_Book,Carmichael_Book}
\begin{equation}\label{ME1}
    \frac{d\hat{\rho}}{dz} =-\mathrm{i}g\hat{S}\hat{\rho} +\gamma_a\mathcal{L}[\hat{a}]\hat{\rho}+\gamma_b\mathcal{L}[\hat{b}]\hat{\rho},
\end{equation}
where $\hat{S}\hat{\rho}=[\hat{a}\hat{b}^\dagger+\hat{a}^\dagger\hat{b},\hat{\rho}]$ determines the interaction of the two fields or light modes, while the second term of the right-hand side is the dissipation superoperator or quantum jump superoperator, with
$\mathcal{L}\left[\hat{c}\right]\hat{\rho}:=2\hat{c}\hat{\rho}\hat{c}^\dagger-\left(\hat{c}^\dagger\hat{c}\hat{\rho}+\hat{\rho}\hat{c}^\dagger\hat{c}\right)$, being $\hat{c}=\hat{a},\hat{b}$. Here, $\hat{a}$ ($\hat{a}^\dagger$) and $\hat{b}$ ($\hat{b}^\dagger$) are the usual annihilation (creation) operators, and $g$ denotes the coupling strength between the two field bosonic modes.

In order to solve the Lindblad master equation, it is important to note that the superoperator $2\hat{c}\hat{\rho}\hat{c}^\dagger$ prevents to rewrite or reinterpret such an equation in terms of a von Neumann-like equation, such that: $d\hat{\varrho}/dz =\mathrm{i}(\hat{H}_{\mathrm{eff}}\hat{\varrho}-\hat{\varrho}\hat{H}_{\mathrm{eff}}^\dagger)$. Despite this, by defining the superoperators $\hat{J}_c\hat{\rho}:=2\hat{c}\hat{\rho}\hat{c}$ and $\hat{L}_c\hat{\rho}:=\hat{c}^\dagger\hat{c}\hat{\rho}+\hat{\rho}\hat{c}^\dagger\hat{c}$, it is not difficult to show that $[(\hat{J}_a+\hat{J}_b),\hat{S}]\hat{\rho}=0$, and $[\hat{J}_c,\hat{L}_c]\hat{\rho}=2\hat{J}_c\hat{\rho}$. This simple, but decisive finding, allows us to perform the transformation
\begin{equation}\label{Trho}
    \hat{\rho}=\exp \left[-\chi\left(\hat{J}_a+\hat{J}_b\right) \right]\hat{\varrho},
\end{equation}
to obtain
\begin{equation}
\begin{split}
    \frac{d\hat{\varrho}}{dz}=&-\mathrm{i}g\hat{S}\hat{\varrho}+\gamma_a\left[\left(1-2\chi\right)\hat{J}_a-\hat{L}_a\right]\hat{\varrho}\\&+\gamma_b\left[\left(1-2\chi\right)\hat{J}_b-\hat{L}_b\right]\hat{\varrho},
\end{split}
\end{equation}
and by setting $\chi=1/2$, the above equation reads
\begin{equation}
\label{rho_J}
\frac{d\hat{\varrho}}{dz}=-\mathrm{i}\left(\hat{H}_{\mathrm{eff}}\hat{\varrho}-\hat{\varrho}\hat{H}_{\mathrm{eff}}^\dagger\right),
\end{equation}
where $\hat{H}_{\mathrm{eff}}=-\mathrm{i}\gamma_a\hat{a}^\dagger\hat{a}-\mathrm{i}\gamma_b\hat{b}^\dagger\hat{b}+g\left(\hat{a}{b}^\dagger+\hat{a}^\dagger\hat{b}\right)$ is an effective non-Hermitian Hamiltonian.\\
We can note that the density matrices $\hat{\rho}$ or $\hat{\varrho}$, governed by \eqref{ME1} or \eqref{rho_J}, respectively, will evolve in two different ways~\cite{Minganti_2019,Minganti_2020,Arkhipov_2020}. A sudden change in the state of the system due to the term $(\hat{J}_a+\hat{J}_b)\hat{\rho}$ is considered in the Lindblad master equation as an average of many quantum trajectories after many experimental realizations or numerical stochastic simulations~\cite{Carmichael_Book,Dum_1992,Molmer_1993}; on the other hand, the evolution of the density matrix undergoes a passive dissipation processes as the quantum jump superoperators are suppressed. Nevertheless, we show that it is possible to establish a direct equivalence between both approaches, through the transformation $\hat{\rho} = e^{-(\hat{J}_a+\hat{J}_b)/2}\hat{\varrho}$. This is one of the main contributions of this work.

Now, from \eqref{rho_J}, we can recognize that the evolution of the density matrix is governed by a von Neumann-like equation, which in turn can be interpreted as two Schrödinger equations due to the non-Hermitian nature of $\hat{H}_{\mathrm{eff}}$, i.e., $\mathrm{i}\frac{d}{dt}\ket{\psi_{R}}=\hat{H}_{\mathrm{eff}}\ket{\psi_{R}}$ and $-\mathrm{i}\frac{d}{dt} \bra{\psi_L}=\bra{\psi_L}\hat{H}_{\mathrm{eff}}^\dagger$, with $\hat{\varrho} = \ket{\psi_R}\bra{\psi_L}$. In order to find a diagonal form of $\hat{H}_{\mathrm{eff}}$ (or $\hat{H}_{\mathrm{eff}}^\dagger$), it is useful and instructive to rewrite it as
\begin{equation}
\begin{split}\label{H_eff}
    \hat{H}_{\mathrm{eff}}=&-\mathrm{i}\left[\frac{\gamma}{2}\left(\hat{a}^\dagger\hat{a}+\hat{b}^\dagger\hat{b}\right)+\frac{\Delta}{2}\left(\hat{b}^\dagger\hat{b}-\hat{a}^\dagger\hat{a}\right)\right]    
    \\&    +g\left(\hat{a}\hat{b}^\dagger+\hat{a}^\dagger\hat{b}\right)
\end{split}
\end{equation}
where $\gamma=\gamma_a+\gamma_b$ and $\Delta=\gamma_b-\gamma_a$; this allows us to notice that 
\begin{equation}
\begin{split}
\hat{N} &= \hat{a}^\dagger\hat{a}+\hat{b}^\dagger\hat{b},\\
\hat{J}_x&=\frac{1}{2}\left(\hat{a}\hat{b}^\dagger+\hat{a}^\dagger\hat{b}\right),\\
\hat{J}_y&=\frac{\mathrm{i}}{2}\left(\hat{a}^\dagger\hat{b}-\hat{a}\hat{b}^\dagger\right),\\
\hat{J}_z&=\frac{1}{2}\left(\hat{b}^\dagger\hat{b}-\hat{a}^\dagger\hat{a}\right),
\end{split}
\end{equation}
satisfy the commutation rules $[\hat{J}_j,\hat{J}_k]=\mathrm{i}\epsilon_{jkl}\hat{J}_l$ (right-hand rule), and $[\hat{N},\hat{J}_j]=0$ (with $j=x,y,z$). We apply the non-unitary transformation $\mathcal{\hat{R}}= e^{\eta\hat{J}_y}$, namely $\hat{\mathcal{H}}=\mathcal{\hat{R}}^{-1}\hat{H}_{\mathrm{eff}}\mathcal{\hat{R}}$, and use the following identities
\begin{equation}
\begin{split}
\hat{\mathcal{R}}^{-1}\hat{J}_z\hat{\mathcal{R}}&=\cosh(\eta)\hat{J}_z-\mathrm{i}\sinh(\eta)\hat{J}_x,\\
\hat{\mathcal{R}}^{-1}\hat{J}_x\hat{\mathcal{R}}&=\cosh(\eta)\hat{J}_x+\mathrm{i}\sinh(\eta)\hat{J}_z,
\end{split}
\end{equation}
the get the transformed Hamiltonian
\begin{equation}
\begin{split}
\hat{\mathcal{H}}=&-\mathrm{i}\frac{\gamma}{2}\hat{N}+\left[2g\cosh(\eta)-\Delta\sinh(\eta)\right]\hat{J}_x\\&+\mathrm{i}\left[2g\sinh(\eta)-\Delta\cosh(\eta)\right]\hat{J}_z.
\end{split}
\end{equation}
Note that although $\hat{\mathcal{R}}$ is a non-unitary transformation, it preserves the commutation rules, but not the Hermiticity~\cite{Balian_1969}. Therefore, from the above equation, with $\tanh(\eta)=2g/\Delta$, the diagonal form of $\hat{\mathcal{H}}$ is 
\begin{equation}\label{H_diag}
\hat{\mathcal{H}}_{\mathrm{diag}}=\frac{1}{2}\begin{cases} -\mathrm{i}\gamma\hat{N}+\omega_{\mathrm{I}}\left(\hat{b}^\dagger\hat{b}-\hat{a}^\dagger\hat{a}\right),&\mbox{if}\quad\Delta\leq 2g,\\
-\mathrm{i}\left[\gamma\hat{N}+\omega_{\mathrm{II}}\left(\hat{b}^\dagger\hat{b}-\hat{a}^\dagger\hat{a}\right)\right],&\mbox{if}\quad\Delta\geq2g,\\
\end{cases}
\end{equation}
and whose eigenvalues in the transformed diagonal basis are 
\begin{equation}\label{eigenvalores}
\lambda_{jk}=\frac{1}{2}
\begin{cases}
-\mathrm{i}\gamma(j+k)+\omega_\mathrm{I}(k-j),&\mbox{if}\quad\Delta\leq 2g,\\
-\mathrm{i}\left[\gamma(j+k)+\omega_\mathrm{II}(k-j)\right],&\mbox{if}\quad\Delta\geq2g,\\
\end{cases}
\end{equation}
where $\omega_{\mathrm{I}}=\sqrt{4g^2-\Delta^2}$, and $\omega_{\mathrm{II}}=\sqrt{\Delta^2-4g^2}$.\\
It is important to point out that at the transition point, $\Delta = 2g$, both eigenvalues and eigenvectors coalesce, and such singularities are called EPs~\cite{Kato}. Furthermore, the aforementioned singularities are associated with symmetry breaking for $\mathcal{PT}$-symmetric Hamiltonians, i.e., the eigenvalues are real before the coalescence point, and complex after that point~\cite{Bender_1998,Heiss_2012}. As can be seen from \eqref{eigenvalores}, the eigenvalues have both real and imaginary part before the coalescence point; however, it has been shown that such system have the same dynamics as the former, and are often called quasi-$\mathcal{PT}$-symmetric systems~\cite{Guo_2009,Ornigotti_2014}.  

On the other hand, although $\hat{\mathcal{H}}_{\mathrm{diag}}$ and $\hat{H}_{\mathrm{eff}}$ (or $\hat{\mathcal{H}}_{\mathrm{diag}}^\dagger$ and $\hat{H}_{\mathrm{eff}}^\dagger$) share eigenvalues, this is not the case for the eigenvectors. Multiplying the left-hand side of $\hat{\mathcal{H}}_{\mathrm{diag}}\ket{j,k} = \lambda_{jk}\ket{j,k}$ by $\hat{\mathcal{R}}$ (or $\hat{\mathcal{R}}^{-1}$), one can recognize that the eigenvalues of $\hat{H}_{\mathrm{eff}}$ or $\hat{H}_{\mathrm{eff}}^\dagger$ are determined by $\hat{H}_{\mathrm{eff}}\hat{\mathcal{R}}\ket{j,k}=\lambda_{j,k}\hat{\mathcal{R}}\ket{j,k}$ or $\bra{j,k}\hat{\mathcal{R}}^{-1}\hat{H}_{\mathrm{eff}}^\dagger=\bra{j,k}\hat{\mathcal{R}}^{-1}\lambda_{jk}^*$, and in turn the eigenvectors by
\begin{equation}\label{eta_LR}
\begin{split}
\ket{\eta_R} &= \hat{\mathcal{R}}\ket{j,k},\\
\bra{\eta_L} &= \bra{j,k}\hat{\mathcal{R}}^{-1}.
\end{split}
\end{equation}
This is an intrinsic consequence of the non-Hermitian nature of $\hat{H}_{\mathrm{eff}}$; its eigenvectors are not orthogonal, i.e., $\braket{\eta_R|\eta_R}\neq0$ or $\braket{\eta_L|\eta_L}\neq0$. It is known that $\ket{\eta_R}$ and $\bra{\eta_L}$ form a bi-orthogonal system~\cite{Rosas_2018}, such that, $\braket{\eta_L|\eta_R} = \braket{j,k|\hat{\mathcal{R}}^{-1}\hat{\mathcal{R}}|l,m} = \delta_{j,l}\delta_{k,m}$. For example, in the semiclassical approach (non-unitary evolution by dropping the quantum jump superoperators $\hat{J}_c\hat{\rho}$), it has been demonstrated that it is possible to generate high-order photon exceptional points by exciting non-Hermitian waveguide arrangements with coherent states~\cite{Tschernig_2022}. However, the bi-orthogonal system determined by \eqref{eta_LR} generalizes the propagation of this type of states in photonics systems with passive losses, and its possible application to generate EPs of arbitrary order.

Finally, having established $\hat{\mathcal{H}}_{\mathrm{diag}} = \hat{\mathcal{R}}^{-1}\hat{H}_{\mathrm{eff}}\hat{\mathcal{R}}$, by direct integration of the $\mathrm{i}\frac{d}{dt}\ket{\psi_R} = \hat{H}_{\mathrm{eff}}\ket{\psi_R}$, and remembering that $\hat{\rho} = e^{-(\hat{J}_a+\hat{J}_b)/2}\hat{\varrho}$, with $\hat{\varrho} = \ket{\psi_R}\bra{\psi_L}$, the exact solution of the Lindblad master equation \eqref{ME1}, given an initial condition $\hat{\rho}(0)$, is
\begin{equation}\label{rho}
\hat{\rho}(z)=e^{-(\hat{J}_a+\hat{J}_b)/2}\hat{U}(z)e^{(\hat{J}_a+\hat{J}_b)/2}\hat{\rho}(0)\hat{U}^{-1}(z),
\end{equation}
where 
\begin{equation}
    \hat{U}(z)= e^{\eta\hat{J}_y}e^{-\mathrm{i}\hat{\mathcal{H}}_{\mathrm{diag}}z}e^{-\eta\hat{J}_y}
\end{equation}
is the non-unitary evolution operator associated to $\hat{H}_{\mathrm{eff}}$. On account of this, the evolution given an input state is determined by three ingredients. By developing in Taylor series $e^{(\hat{J}_a+\hat{J}_b)/2}$ and applying the powers of $(\hat{J}_a+\hat{J}_b)/2$ to the input  state, we obtain a new state (pure or mixed); this new state will evolve according to $\hat{U}(z)$ to then be transformed as in the first step and finally obtain the output state. In other words, the dynamics of two-coupled fields subject to Markovian conditions is equivalent to two fields or light states crossing a non-Hermitian system with passive losses, provided that the input and output states are transformed according to $e^{\pm(\hat{J}_a+\hat{J}_b)/2}[\hat{\bullet}]$ (see Fig.~\ref{Fig_1}).

In addition, due to the mathematical similarities between paraxial optics and the Schrödinger equation, one of the quintessential platforms for propagating quantum states in non-Hermitian photonic systems at the single-photon level are evanescently coupled waveguides~\cite{ElGanainy_2007,Ruter_2010}. In such systems, the photonic state is affected along the propagation distance due to the balance between gains and losses, giving rise to the concept of $\mathcal{PT}$-symmetry. When only losses (passive losses) are considered, this is known as passive or quasi-$\mathcal{PT}$-symmetry~\cite{Guo_2009,Ornigotti_2014}. Furthermore, under Markovian assumptions, two-photon interference in two evanescently coupled waveguides has recently been demonstrated, where losses are induced through an asymmetric refractive index distribution and modulation along the propagation distance~\cite{Klauck_2019}. However, we have shown that the dynamics of the photon density matrix in such non-Hermitian systems does not require extra modulation along the propagation distance (there are only passive losses, see Fig.~\ref{Fig_1}). Nonetheless, the exchange of energy in non-conservative systems with their surrounding environment, described in terms of \eqref{ME1} or \eqref{rho_J}, is fundamentally different, since the quantum jump superoperators have a great influence on the propagation of non-classical states and a strong impact on the spectral responses of dissipative systems~\cite{Minganti_2019,Arkhipov_2020,Minganti_2020}. Although non-Hermitian $\mathcal{PT}$ systems are the prelude to open quantum systems, a direct equivalence between the two has not been found so far. However, we show that both systems are equivalent as long as the photonic density matrix is transformed via \eqref{Trho}.

\begin{figure}[ht!]
\begin{center}
\includegraphics{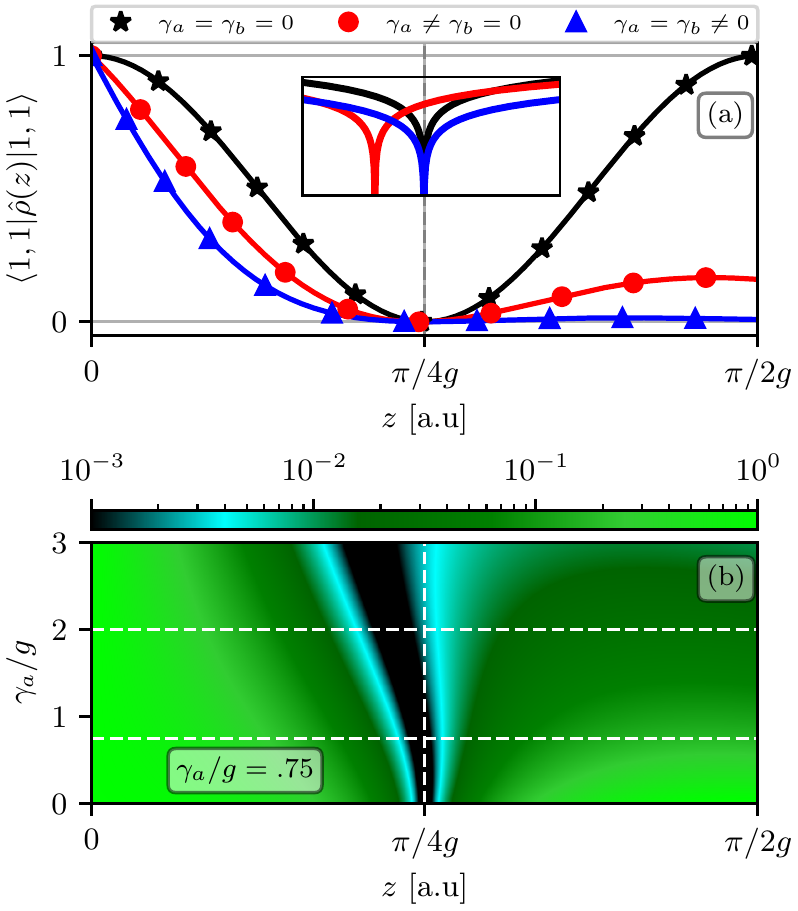}
\caption{(a)~Two photons are launched into two waveguides with passive losses. At $z=\pi/4g$, when both waveguides have the same decay rate or the rate is zero, the coincidence rate is zero (black and blue lines). However, when we consider only passive losses in one of the two waveguides, the minimum value of the coincidence rate undergoes a shift (red line). In panel (b), we plot the coincidence rate as a function of propagation distance and decay rate $\gamma_a/g$ (with $\gamma_b=0$). For $\gamma_a/g=0.75$, we recover the previous case. As $\gamma_a/g$ increases, we observe that the minimum value of the coincidence rate occurs at smaller distances, and at $z=\pi/4g$, the photons tend to anti-bunch. The analytical results are shown as continuous lines, while the numerical simulations are represented by dots, stars, and triangles.}
\label{Fig_2}
\end{center}
\end{figure}
For example, in order to test our analytical results, we investigate two-photon quantum interference in two evanescently coupled waveguides with passive losses. The photonic density matrix $\hat{\rho}(z)$ evolves as a function of propagation distance, and for two indistinguishable photons $\ket{1,1}$, we can compute the coincidence rate at distance $z=z$, taking into account first of all that
\begin{equation}
\begin{split}    e^{(\hat{J}_a+\hat{J}_b)/2}\ket{1,1}\bra{1,1}=& \ket{0,0}\bra{0,0}+\ket{1,1}\bra{1,1}\\&+\ket{0,1}\bra{0,1}+\ket{1,0}\bra{1,0}
\end{split}
\end{equation}
and
\begin{equation}
\begin{split}
    e^{\eta\hat{J}_y}\ket{1,1} &= e^{-\eta\hat{J}_y}\hat{a}^\dagger\hat{b}^\dagger e^{\eta\hat{J}_y}e^{-\eta\hat{J}_y}\ket{0,0}\\ &= \cosh (\eta )\ket{1,1}+\frac{\mathrm{i}}{2}  \sinh (\eta)\left[\ket{0,2}-\ket{2,0}\right]
\end{split}
\end{equation}
and the other equivalent operations involved in \eqref{rho}, such that
\begin{equation}
\braket{1,1|\hat{\rho}(z)|1,1} =e^{-2\gamma z} 
\begin{cases}
\left[\frac{ 4g^2 \cos(\omega_\mathrm{I}z) - \Delta^2}{\omega_\mathrm{I}^2} \right]^{2},~\mbox{if}~\Delta\leq 2g,\\
\left[ \frac{\Delta^2 - 4g^2 \cosh(\omega_{\mathrm{II}} z)}{\omega^{2}_{\mathrm{II}}}\right]^{2},~\mbox{if}~\Delta\geq 2g.
\end{cases}
\end{equation}
In Fig.~\ref{Fig_2}, we show two complementary results associated with the coincidence rate as a function of propagation distance and the loss rates $\gamma_a$ and $\gamma_b$, respectively.
\begin{itemize}
    \item (a) For both unitary and non-unitary evolutions, i.e., $\gamma_a=\gamma_b=0$ and $\gamma_a=\gamma_b\neq0$, respectively, the bunching occurs at $z=\pi/4g$. However, when the losses are asymmetric, i.e., for example, $\gamma_a/g=.75$ and $\gamma_b=0$, the bunching occurs at a smaller distance than in the previous case. This fact has already been reported experimentally in the framework of the Hong-Ou-Mandel dip in~\cite{Klauck_2019}. However, the photonic system exhibiting $\mathcal{PT}$ symmetry is different. In our case, no extra bending modulation is necessary to generate Markovian losses in either of the two waveguides. We show that the Markovian signature is inscribed outside the photonic system exhibiting $\mathcal{PT}$ symmetry, as it suffices to apply the transformation $e^{\pm(\hat{J}_a+\hat{J}_b)/2}[\hat{\bullet}]$ to the input and output channels. 
    \item (b) As we increase the losses asymmetrically ($\gamma_b=0$), we can notice that photon bunching occurs at distances smaller and smaller than $z=\pi/4g$. On the other hand, at the $\mathcal{PT}$ symmetry phase transition point ($\gamma_a/g = 2$), we can observe anti-bunching, i.e., the coincidence rate increases and the photons tend to exit separately from each of the channels. In \cite{Longhi_2020}, a unitary transformation was proposed to verify this fact, which involves rotating the mode basis of the input and output.
\end{itemize}

In this work, we solve \eqref{ME1} for any set of parameters $\gamma_j$ and $g$ (with $j=a,b$, respectively). This solution determines the evolution of the photonic density matrix $\hat{\rho}(z)$, given by \eqref{rho}, as a function of the propagation distance or time. As a consequence of this solution, we can reinterpret the dynamics of non-classical states in systems exhibiting $\mathcal{PT}$ symmetry. For example, our system exhibiting $\mathcal{PT}$ symmetry requires only passive losses in each of the waveguides (or directional coupler) as long as the initial and final conditions are transformed by $e^{\pm(\hat{J}_a+\hat{J}_b)/2}[\hat{\bullet}]$ (see Fig.~\ref{Fig_1}). This means, and it is important to re-emphasize, that non-Hermitian systems are not only a prelude to systems with Markovian losses but also that it is possible to establish a direct equivalence relation without omitting the quantum jump superoperators, at least for two quantized fields subject to Markovian losses. We believe that this result will allow greater control and experimental feasibility when propagating non-classical states in systems exhibiting some kind of $\mathcal{PT}$ symmetry.
%
\end{document}